\documentstyle[aps,manuscript]{revtex}  

\newcommand{\grad}[0]{\mbox{\boldmath $\nabla$}}
\newcommand{\bH}[0]{\mbox{\boldmath $H$}}
\newcommand{\cH}[0]{ {\cal H} }
\newcommand{\br}[0]{\mbox{\boldmath $r$}}

\begin{document}

\title{Dynamic and Static Properties of\\
 the Randomly Pinned Planar Flux Array}

\author{G. George Batrouni}
\address{Thinking Machines Corporation\\
245 First Street\\
Cambridge, Massachusetts  02142}

\author{Terence Hwa\cite{addr}}
\address{School of Natural Sciences\\
Institute for Advanced Study\\
Princeton, New Jersey  08540}

\date{\today}
\maketitle
\widetext
\begin{abstract}
We report the results of large scale numerical studies of the dynamic
and static properties of the random phase model of the randomly pinned
flux lines confined in a plane.
  From the onset of nonlinear IV characteristics, we identify a
vortex glass transition at the theoretically anticipated temperature.
However, the vortex glass phase itself is much {\it less glassy} than
expected. No signature of the glass transition has been detected in the
static quantities measured.
\end{abstract}


\narrowtext

Flux pinning is crucial to the performance of high-$T_c$ superconductors in
a strong magnetic field~\cite{rmp}. It has been conjectured~\cite{mpaf,ffh}
that flux lines in a superconductor with random, point-like impurities may
form a glass, the vortex glass, in which the flux line configurations are
frozen by the defects at low  temperatures.  It is
argued~\cite{ffh,creep} that {\it if} a glass phase exists, then the sluggish
glassy dynamics of the flux lines suppresses dissipation, making the system
a {\it true} superconductor.  Unfortunately,
evidence supporting the vortex glass hypothesis is still inconclusive: The
phase transition seen experimentally by Koch {\it et al}~\cite{koch} and by
Gammel {\it et al}~\cite{gammel} may be due to pinning by {\it correlated}
defects~\cite{nv} such as twin planes and/or screw dislocations. Existing
numerical simulations which support the existence of the vortex glass are
thus far restricted to models with zero external magnetic field and/or an
infinite London penetration length~\cite{num}. However, the glass phase
seems to be unstable to thermal fluctuations in more realistic models with
a finite London penetration length~\cite{young}.

Without solid evidence from experimental and numerical studies, major
support for the vortex glass hypothesis comes from a number of analytic
studies~\cite{mpaf,creep,yedidia,ledou}, which are however
 restricted to ``elastic
systems" where topological defects are forbidden to nucleate. While the
validity of the elastic approximation for the bulk system in $2+1$
dimensions is a subject of intense current research, the elastic model is
well justified to describe flux lines confined in a plane ($1+1$
dimensions)~\cite{mpaf,nat,toner} or vortex lines in planar Josephson
junctions~\cite{vinokur}.  Thus far, the $1+1$ dimensional flux array is
the {\it only} system for which a vortex glass phase has been demonstrated
analytically; as such, it is one of the very few pieces of ``solid" support
on which the vortex glass hypothesis for bulk superconductors is based.
Because of this significance, the static and dynamic properties of the
$1+1$ dimensional
system have been investigated analytically by many groups
in the past several
years~\cite{mpaf,yedidia,ledou,nat,toner,vinokur,nat2,shapir,sudbo,
comments,balents,tsvelik,korshunov,hf}.
These studies led to a number of rather  different
predictions, none of which has been tested experimentally or numerically.

In this paper, we report the results of detailed numerical studies of the
random phase model~\cite{mpaf,nat,vinokur,hnv} of the planar flux array
using the Connection Machine CM5.  We observe an apparent vortex glass
transition at a finite temperature, $T_g$, below which the IV
characteristics are nonlinear and described by power laws, as anticipated
by Refs.~\cite{nat} and \cite{shapir}.  However, the universal constant
found is significantly smaller than that predicted in Ref.~\cite{shapir}.
Static quantities such as the equal-time correlation function are also
measured. They exhibit no detectable signs of the glass transition.

We consider an array of flux lines confined to the $(x,z)$-plane, and
directed in the $z$-direction by an applied magnetic field $\bH = H_0
\hat z$~\cite{hnv}. Line-line
repulsion is modeled by linear elasticity, with an elastic constant
$\kappa = (d\rho/dH_0)^{-1}$, where $\rho$ is the average line density.
Random
point-like pinning centers in the plane are described by an uncorrelated
Gaussian random potential with a variance $\Delta_0$.  A direct simulation
of the flux array is deemed difficult because an interline spacing of
$\rho^{-1} \sim 10$ grid points and a system size of $ \gg 10 \rho^{-1}$ is
needed to probe the asymptotic behavior.  Instead we will study the
random phase model,
\begin{equation}
\cH = \int d^2\br \left\{ \frac{\kappa}{2} (\grad u)^2 - \lambda
\cos[2\pi u(\br) -\beta(\br)] \right\}, \label{H2}
\end{equation}
which is generally believed~\cite{mpaf,nat,hnv} to describe the statistical
mechanics of the random flux array at length scales large compare to the
inter-line spacing $\rho^{-1}$.  In Eq.~(\ref{H2}), $\br = \{x,z\}$,
$u(\br)$ is a displacement-like field, the cosine term captures the
discreteness of the flux lines crucial to pinning, $\beta(\br)$ is
a random phase uniformly distributed in the interval $[0,2\pi]$,
and $\lambda = 2\sqrt{\Delta_0}\rho$ characterizes the strength of the
pinning potential.  A derivation of this model starting from a system of
directed flux lines can be found in Ref.~\cite{hnv}.

Renormalization group (RG) studies of Eq.~(\ref{H2})~\cite{cardy,sr} found a
glass transition at $T_g = \kappa/\pi$ for small $\lambda$.  For $T>T_g$,
randomness (the cosine term) is irrelevant, and the system is
asymptotically described by Eq.~(\ref{H2}) with $\lambda=0$.  But for $T<
T_g$, the system is described by a line of fixed points.  A number of
properties of the glass phase close to $T_g$ have been obtained from the RG
studies. For instance, the static correlation function is predicted to
diverge with a logarithmic anomaly~\cite{sr},
\begin{equation}
C(\br) \equiv \overline{\langle( u(\br) - u(0) )^2\rangle}
=  A C_0(\br) + B \tau^2 [ \ln|\br|]^2,   \label{corr}
\end{equation}
where $\tau \equiv 1-T/T_g$, $C_0(\br) = T/(\pi\kappa) \ln |\br|$ is
the correlation function with $\lambda = 0$, $A$ is a nonuniversal
constant,  and $B =2/\pi^2 + O(\tau)$ is universal~\cite{hf}.  Overbar and
$\langle\ldots\rangle$ denote disorder and thermal averages respectively.

Dynamic RG methods were used to extract the glassy dynamics
of the model (\ref{H2}) close to $T_g$~\cite{shapir,dxy}.
In particular, under a small driving
force $F$, the Langevin dynamics
\begin{equation}
\partial_t u = -\frac{\delta \cH}{\delta u}
+ F + \eta, \label{eom}
\end{equation}
where $\eta(\br,t)$ is a white noise with
$\langle \eta(\br,t) \eta(0,0) \rangle
= 2 T \delta^2(\br) \delta(t)$,
is expected to yield a nontrivial response,
$v(F)\equiv \overline{\langle\partial_t u\rangle}$, with
\begin{eqnarray}
&v(F) \propto |\tau|^\zeta F, \qquad \qquad \qquad \qquad
\qquad &{\rm for} \qquad T>T_g, \label{v1}\\
&v(F) \propto |\tau|^\zeta F^\alpha, \qquad \alpha = 1 + \zeta\tau
\qquad &{\rm for}\qquad T< T_g, \label{v2}
\end{eqnarray}
with the universal constant $\zeta \approx 1.78 + O(\tau)$.  $F$ describes
the effect of a Lorentz force resulting from an electric current
perpendicular to the $(x,z)$-plane.  $v(F)$, the ``IV characteristics'', is
proportional to the EMF generated from the in-plane motion of the flux
lines when pushed by the Lorentz force~\cite{toner}.  A zero DC resistivity
($\lim_{F\to 0} v(F)/F \to 0$) and hence true superconductivity is expected
in the vortex glass phase according to Eq.~(\ref{v2}).

Within the RG framework, these predictions are valid in the asymptotic
limits of the system size $L\to \infty$ and $F \to 0$.
Finite size effects are derived using the length dependence of the
dimensionless coupling constant $g(L)$,
\begin{equation}
g(L) = \frac{g(1) L^{2\tau}}{ 1 + Cg(1) [L^{2\tau}-1]/(2\tau)},\label{dg}
\end{equation}
which characterizes the effective strength of disorder at scale $L$, and is
 obtained from the RG recursion relation~\cite{hf,sr}.
In Eq.~(\ref{dg}), $g(1) =\pi\lambda^2/(2T^2)$ is the bare coupling
constant, and the nonuniversal coefficient $C\approx 1$ controls the crossover
length for the model (\ref{H2}).   Finite size effects in the
correlation function shall be estimated by replacing $\tau$ in
Eq.~(\ref{corr}) by $(2/C)g(L)$~\cite{note1}.

When $F\ne 0$, the equilibration time  is $t_{\rm eq} \approx
1/v(F)$, since for $v(F) t \gg 1$, $\langle u \rangle
\approx v(F) t \gg 1$, which averages out the pinning effect of the
cosine~\cite{shapir}.  But during the time $t_{\rm eq}$, only regions of
size $L_{\rm eff} \approx t_{\rm eq}^{1/z}$ ($z=2\alpha$~\cite{shapir}) can
be equilibrated. Hence the effective system size at a finite drive is
$L_{\rm eff} \approx F^{-1/2}$.  While the effect of a large $F$ involves
nonequilibrium dynamics~\cite{shapir} and is complicated, at the level of
linear response, we shall again estimate the effect of a finite $F$ by
replacing the coefficient $|\tau|^\zeta$ in Eq.~(\ref{v2}) by
$[(2/C)g(F^{-1/2})]^\zeta$.

Although all existing analytic studies agree that $T_g=\kappa/\pi$,
some results of the RG analysis disagree with results by other
methods. For instance, Toner~\cite{toner} argued that the form of the IV,
for $T<T_g$, should be
\begin{equation}
v(F) \sim F \ e^{-c\sqrt{\log (F_0/F)}}.  \label{v3}
\end{equation}
Also, Balents and Kardar~\cite{balents} obtained, at $T=T_g$,
 a much more strongly divergent correlation function
\begin{equation}
C(\br) \sim |\br|, \label{corr2}
\end{equation}
using a Bethe ansatz solution of the random flux array.
On the other hand, a similar study by Tsvelik~\cite{tsvelik}
yielded $C(\br) \propto \ln|\br|$ when a different order of thermodynamic
limits was taken.   One
criticism of the RG analysis is that the possibility of ``replica symmetry
breaking", often associated with the existence of meta-stable states, has
not been considered.  A recent variational ansatz with hierarchical
replica symmetry breaking scheme~\cite{ledou,korshunov} yields a
correlation function even less divergent than Eq.~(\ref{corr}),
\begin{equation}
C(\br) = \frac{T_g}{\pi\kappa} \ln |\br| \label{corr3}
\end{equation}
for {\it all} $T < T_g$ and $|\br| \gg 1$.
However, an earlier attempt~\cite{yedidia} yielded Eq.~(\ref{corr2}).

In view of the myriads of predictions for this system, we
perform a large scale numerical study. As we
shall be interested in the dynamics, we directly simulate the Langevin
equation (\ref{eom}) of the model (\ref{H2}) on an $L\times L$ square
lattice with periodic boundary conditions in both the $x$ and $z$
directions. To do this numerical integration, we approximate the time
derivative in Eq.~(\ref{eom}) by a finite difference, and use a two step,
single noise Runge-Kutta integration algorithm~\cite{helfand,ggb1,ggb2}.
This algorithm yields results accurate to $O(dt^2)$, where $dt$ is
the the discrete time step. We ensure that the systematic $dt^2$ errors are
as predicted and under control by doing runs at various $dt$'s (with $1024$
realizations each) and various temperatures~\cite{ggb2}. The simulations
reported below are for runs with an optimized $dt = 0.2$, and sizes $L=64,
128$. The parameters are fixed at $\kappa = 1.0$ and $\lambda = 0.15$.
The IV curves are obtained by varying $F$ in the range $0.005$ to $0.1$,
and the glass transition probed by varying $T$ from $0.5$ down to $0.25$.
The number of realizations over which we averaged ranged from 32 ($L=128$)
to 1024 ($L=64$) and is specified where relevant. Note that the range
of parameters are chosen such that $g(1) > g(L) > g(\infty)$ according to
Eq.~(\ref{dg}). Thus the effective strength of disorder should appear
{\it stronger} than its asymptotic value due to the finite size effect.
However, this effect is not expected to be large, since the crossover
length, $L_c$, defined by equating the two terms of the denominator of
Eq.~(\ref{dg}), is only $L_c = e^{\frac{1}{Cg(1)}} \approx 15$ at its largest
(when $\tau=0$)~\cite{ggb2}.

We start from random
initial conditions in $u$. Depending on the sizes and temperatures of the
systems, we typically use $80,000$ to $200,000$ time steps to equilibrate,
and then $60,000$ to $150,000$ steps to perform measurements.  ``Thermal
averages" are carried out by time-averaging over the measurement period for
each realization.  In all
cases, equilibration is assured by monitoring the susceptibility,
$\chi = \langle(\int d^2\br\
\partial_x u)^2 \rangle/(TL_xL_z)$, whose mean is
$1/\kappa$, at all $T$, within statistical and $dt^2$ errors~\cite{hf,ggb2}.
 As extra
precaution, we did some very long runs ($10^6$ steps, $160$ realizations)
for the $L=64$ system at $T=0.285$. All measurements there agree
with the shorter runs.


We begin by examining the IV curves $v(F)$.  Figure 1 shows the IV curves
for a $64\times 64$ lattice for various temperatures on a log-log plot.
(Note that the vertical axis is divided by $F$ to emphasize the nonlinear
part of $v(F)$.)  According to Eqs.~(\ref{v1}) and (\ref{v2}), we expect
linear IV curves only above $T_g$. From Fig. 1, we see that the data for
$T=0.4$ satisfy a linear relation~\cite{note}, while for $T = 0.3$ and
below, they are nonlinear and well described by a power law as in
Eq.~(\ref{v2}).  The straightness of the data strongly biases against the
prediction (\ref{v3}).  The $L=128$ lattice yields similar results.  Fitting
Eq.~(\ref{v2}) to the IV curves, we can obtain the critical temperature,
$T_g$, and the coefficient $\zeta$.  From Fig.~2 we see that the exponent
$\alpha$ indeed varies linearly with the reduced temperature $\tau$  up to
at least $\tau \approx 0.21$ (or $T = 0.25$). Chi-squared fits to our IV
data for $T= 0.25, 0.27, 0.285$ and $0.3$ ($L=64$) and $T=0.25, 0.275$ and
$0.3$ ($L=128$), give $T_g=0.32\pm 0.01$ and $\zeta=0.15 \pm 0.02$ for the
$L=64$ system, and $T_g=0.34\pm 0.01$ with $\zeta=0.15 \pm 0.02$ for the
$L=128$ system.  The close agreement between the two systems indicates that
our result is not limited by the system size $L$.
 Our value for $T_g$ agrees with the
predicted value of $1/\pi \approx 0.318$, but the predicted value for the
universal coefficient $\zeta$ is an order of magnitude too large. For
comparison, we indicate the predicted slope of the IV curve at $T=0.25$ as
the dashed line in Fig.~1, and the predicted $\alpha(\tau)$ as the dashed
line in Fig.~2.  Clearly, neither can be accommodated by our numerical
results.

One may attempt to attribute the discrepancy between the numerics and the
RG prediction to crossover effects, since even the smallest value of $F$
used corresponds to only $L_{\rm eff} \approx 15$.  However, since $v(F)
\propto [g(F^{-1/2})]^\zeta F^\alpha$ and $g(F^{-1/2})$ {\it increases}
with $F$ for the range of parameters used, one expects the effective
values of $\alpha$ observed to be somewhat {\it larger} than the expected
asymptotic values~\cite{ggb2}.  It is therefore
not possible to explain the observed discrepancy by such crossover effects.
In fact, it is possible (although unlikely) that the observed
nonlinear IV's are the pre-asymptotics of some {\it linear} IV.
A study at smaller
$F$'s is desirable but difficult numerically, since a much longer measurement
period will be needed to obtain $v(F)$ accurately for each $F$.

Next we consider the static correlation function, $C({\bf r})$, for which
we have much better statistics and smaller finite size effects since $F=0$.
In comparing our data for $C({\bf r})$ with the predicted forms,
Eqs.~(\ref{corr}), (\ref{corr2}), and (\ref{corr3}), we do not compare
directly with $\ln|{\bf r}|$ since that is the form for the quadratic
theory in the continuum. Instead we compare with the correlation function
$C_0(\br)$ of the quadratic {\it lattice} model calculated exactly for
$dt=0.2$.  Figure 3
shows the measured values of $C({\bf r})$ at $T =0.3, 0.285$ and $0.25$
for the $L=64$ system each averaged over $1024$
realizations. (The error bars are much smaller than the points and are not
shown.)  The data are well described by $C(\br) = A C_0(\br)$ with $A
\approx 1.09$ for all $T$, and we find $A \to 1$ as $dt \to 0$
for all $\br$~\cite{ggb2}.
Our results are significantly different from  Eq.~(\ref{corr}),
whose prediction for $C(\br)$
at $T=0.25$ is plotted as the dashed line in Fig.~3 and should be easily
detectable, unless the universal coefficient $B$ is much (two
orders of magnitude) smaller than predicted.
Again, it will be difficult to attribute the discrepancies to finite size
effects which would have made $C(\br)$ appear slightly
{\it larger} since $g(1) > g(\infty)$~\cite{ggb2}.
Our data are also in disagreement with Eqs.~(\ref{corr2}) and
(\ref{corr3}). However, the crossover length to Eq.~(\ref{corr3})
is expected to be long [$O(e^{1/\tau})$] for small $\tau$~\cite{ledou}. Here,
simulations at even lower temperatures are needed to make an unequivocal test.

Finally we examined the variance of the susceptibility $\chi$, which is
expected to be nonzero and universal in the glass phase~\cite{hf}.
However, our numerical results yield a variance that is zero within
statistical error above and below $T_g$~\cite{ggb2}.  It is again in clear
disagreement with the RG prediction and in line with the very small effect of
disorder observed in all of the other measurements.

To summarize, our numerical simulation of the random phase model of the
randomly pinned planar flux array leads to some rather surprising results:
Although the onset of nonlinear IV characteristics is observed at the expected
glass transition temperature, the low temperature phase is found to be much
{\it less} glassy than that predicted by the RG analysis.
Both the disorder-averaged correlation function and the
normalized variance of susceptibility variations are, within statistical
errors, indistinguishable from those of the pure system.  Our numerical
results clearly rule out the RG predictions --- either the universal
constants $\zeta$ and $B$ are respectively at least
one and two orders of magnitude smaller, or the crossover scale is at least
two orders of magnitude longer. Other possible scenaria include:
 (1)  We are prevented from reaching the asymptotics
of some {\it unknown} fixed point (such as those proposed in
Refs.~\cite{ledou,balents,tsvelik,korshunov}) by a very strong
crossover effect. (2) The phase transition observed is purely dynamic and does
not show up in the statics. The source of discrepancies
between theory and simulation is very puzzling and warrants more detailed
theoretical and numerical investigation.

We wish to thank   L.~Balents, D. Cule,
D.~S.~Fisher, P.~Le~Doussal, Y.~Shapir, and Y.-C.~Tsai for discussions.
This work is supported in part by the US Department of Energy Contract
No.~DE-FG02-90ER40542.


\begin{figure}
\vspace{1.5truein}
\caption{The nonlinear part of the IV curves for $L=64$
at $T = 0.4, 0.3, 0.25$. The solid lines are best straight line fits. The
dashed line is Eq.~(5) at $T=0.25$.}
\end{figure}

\begin{figure}
\vspace{1.5truein}
\caption{The exponent $\alpha$ for $L=64$ as a function of
the reduced temperature.  The dashed line is $\alpha = 1 + 1.78\tau$.}
\end{figure}

\begin{figure}
\vspace{1.5truein}
\caption{The correlation function for $L=64$ at $T=0.3, 0.285, 0.25$ with
1024 realizations each.  The solid lines are $C(r) = 1.09 C_0(r)$.  The
dashed line is Eq.~(2) at $T=0.25$ and $A=1$.}
\end{figure}


\begin{references}
\bibitem{note} Due to the large finite size effect close to
$T_g=\kappa/\pi$ (see Eq.~(\ref{dg})),
it is difficult to determine the true temperature
dependence of the slope of $v(F)$ to test Eq.~(\ref{v1}).


\vfill


\pagebreak
\end{references}
\end{document}